\documentclass[conference, a4paper, draftclsnofoot, onecolumn]{IEEEtran}
\usepackage[left=1.57cm,right=1.57cm,top=2.5cm,bottom=2.5cm]{geometry}
\usepackage{cite}
\usepackage{amsmath,amsfonts,amssymb,amsthm} % per l'ambiente matematico
\usepackage{algorithmic}
\usepackage{graphicx}
\usepackage{textcomp}
\usepackage{xcolor}
\usepackage{dcolumn}% Align table columns on decimal point
\usepackage{bm}
\usepackage{graphicx}
\usepackage{multirow}
\usepackage{bigstrut}
\usepackage{caption}
\usepackage{subcaption}
\DeclareCaptionFont{mysize}{\fontsize{8}{9.6}\selectfont}
\usepackage[lighttt]{lmodern}
\usepackage[T1]{fontenc} % font encode
\usepackage[utf8]{inputenc} % input encode, per lettere accentate eccetera
\usepackage[english]{babel} % impostazione lingua
\usepackage{booktabs}
\usepackage{xspace}
\usepackage{changepage} % aiuta nell'impaginazione di documenti scritti in twoside
\usepackage{epigraph} % per le citazioni di inizio capitolo, eventualmente va dato un noindent
\usepackage{hologo} % per i loghi di latex, bibtex ecc...
\usepackage{fancyhdr}			% imposta intestazioni
\usepackage{pgfplots}
\usepackage{bigints}
\usepackage{quoting}
\usepackage{epstopdf}
\usepackage{stackengine}
\usepackage[normalem]{ulem}
\usepackage[hidelinks]{hyperref}
\usepackage[capitalize, nameinlink]{cleveref}
\usepackage[export]{adjustbox}
\usepackage{chemformula}
\usepackage{floatrow}
\usepackage{sidecap}
\usepackage{mwe}
\usepackage{overpic}
\usepackage{siunitx}  % per le unità di misura nell'albiente matematico e fuori
\usepackage{orcidlink}
\usepackage{float}
\usepackage{subcaption}
\floatstyle{plaintop}
\restylefloat{table}

\sidecaptionvpos{figure}{t}
\newcommand{\sidecaption}[1]% #1 = label name
{\raisebox{\abovecaptionskip}{\begin{subfigure}[t]{1.6em}
  \caption[singlelinecheck=off]{}% do not center
  \label{#1}
\end{subfigure}}\ignorespaces}
\newsavebox{\twosubbox}
\include{macros}
\usepackage{color}

\graphicspath{{../images/}{../figures}{../fig/}{../img/}{../FIGURES/}{FIGURE/}{images/}{figures}{fig/}{img/}{FIGURES/}{/}}

\newcommand{\putHelvetica}[4]{\put(#1, #2){\fontsize{#3}{#3}\fontfamily{phv}\selectfont #4}}

\def\BibTeX{{\rm B\kern-.05em{\sc i\kern-.025em b}\kern-.08em
    T\kern-.1667em\lower.7ex\hbox{E}\kern-.125emX}}
\begin{document}
\bstctlcite{IEEEexample:BSTcontrol}
%%%%%%%%%%%%%%%%%%%%%%%%%%%%%%%%%%%%%%%%%%%%%% HEADER %%%%%%%%%%%%%%%%%%%%%%%%%%%%%%%%%%%%%%%%%

\title{Modelling and Simulations of Ferroelectric Materials and Ferroelectric-Based Nanoelectronic Devices \\ (Invited Paper)}
\author{
    \IEEEauthorblockN{
        D. Esseni, F. Driussi, D. Lizzit, M. Massarotto, M. Segatto.
    }

    \IEEEauthorblockA{
        DPIA, University of Udine, Udine, Italy
    }
}
\maketitle
\IEEEpubidadjcol

%%%%%%%%%%%%%%%%%%%%%%%%%%%%%%%%%%%%%%%%%%%%%% ABSTRACT %%%%%%%%%%%%%%%%%%%%%%%%%%%%%%%%%%%%%%%%%

\begin{abstract} This paper provides a brief introduction to the phenomenological aspects of the polarization in ferrroelectric materials, and then an analysis of a few selected topics related to the modelling of ferroelectrics. The description of ferroelectric-based devices is quite challenging, particularly because the ferroelectric is frequently stacked with other dielectrics or with a semiconductor, as opposed to being placed between metal electrodes. Predictive modelling of ferroelectric devices is admittedly difficult, and thus the scrutiny and calibration of the models by comparison to sound experimental data is of paramount importance.

\end{abstract}

%%%%%%%%%%%%%%%%%%%%%%%%%%%%%%%%%%%%%%%%%%%%%% KEYWORDS %%%%%%%%%%%%%%%%%%%%%%%%%%%%%%%%%%%%%%%%%


%%%%%%%%%%%%%%%%%%%%%%%%%%%%%%%%%%%%%%%%%%%%%% CONTENT %%%%%%%%%%%%%%%%%%%%%%%%%%%%%%%%%%%%%%%%%

\noindent\textbf{© 2023 IEEE.  Personal use of this material is permitted.  Permission from IEEE must be obtained for all other uses, in any current or future media, including reprinting/republishing this material for advertising or promotional purposes, creating new collective works, for resale or redistribution to servers or lists, or reuse of any copyrighted component of this work in other works.}

\noindent\textbf{The paper has been presented at the International Conference on Simulation of Semiconductor Processes and Devices (SISPAD) in September 2023, Kobe, Japan.}

\section{Introduction}\label{sec:introduction}

The discovery of ferroelectricity as a physical phenomenon dates back to about one century ago \cite{Valasek_PhysRev1921}.
%, and it was first reported for Rochel salts, that not accidentally exhibit also piezoelectric and pyroelectric properties \cite{Fousek_IEEE1994}. 
Ferroelectric materials display a spontaneous electric polarization (over some range of temperature), that can be oriented by an external electric field and lends itself to many possible nanoelectronic applications.
The switchable polarization in perovskite thin films (e.g. PbZr$_{1-x}$Ti$_x$O$_3$ (PZT), SrBi$_2$Ta$_2$O$_9$ (SBT), BaTiO$_3$ \cite{Rabe_Book}), for example, has been exploited in ferroelectric random access memories (FeRAM) since the nineties \cite{Evans_JSSC1989,Takashima_NVM2011}. Many more applications of ferroelectrics to electronic devices have been proposed after the discovery of ferroelectricity in fluorite-type Hf$_{0.5}$Zr$_{0.5}$O$_2$ \cite{Boscke_IEDM2011,Zhou_ActaMat2015} and, more recently, in the wurtzite-type Al$_{1-x}$Sc$_{x}$N materials \cite{Fichtner_JAP2019}, both materials offering a much better compatibility with CMOS processing compared to perovskites.

The exploitation of ferroelectrics in CMOS devices and circuits offers many challenges and opportunities for modelling and simulations at all abstraction levels. First principles studies are needed to clarify the phase transition paths behind the polarization switching. Physically-based, device-level models can address the stabilization and reversal of the polarization in actual devices, where ferroelectrics are stacked with other dielectrics or semiconductors.
%where the ferroelectric is placed adjacent to a dielectric or a semiconductor film, instead of sitting between two metallic electrode. 
Then, compact models are also necessary to fully harness the ferroelectric properties at the circuit level. 
%The exploration of the retention and endurance of the polarization state, as well as the vertical and lateral scalability of ferroelectric materials and related devices are also topics where modelling and simulations can help explore a huge design space.

In this paper, we will be able to touch only a few of the possible topics related to the modelling of ferroelectric materials and devices, and to provide references to deepen and widen the analysis beyond what we could address in this paper, and in the related presentation.

%%%%%%%%%%%%%%%%%%%%%%%%%%%%%%%%%%%%%%%%%%%%%%%%%%%%
\section{Phenomenological Aspects in Ferroelectric Materials}\label{sec:PhenomenologicalAspects}
%%%%%%%%%%%%%%%%%%%%%%%%%%%%%%%%%%%%%%%%%%%%%%%%%%%%
\begin{figure}
	\centering
	\includegraphics[width=4in]{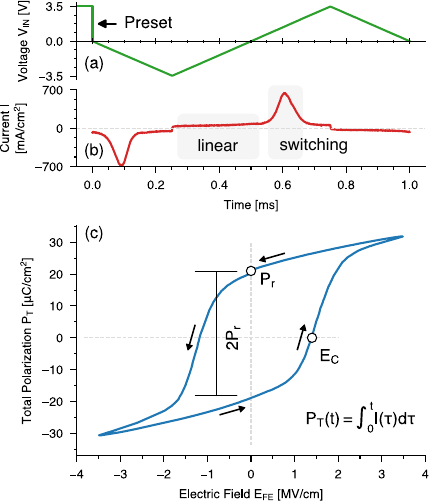}
	\caption[]{Electrical characterization for an MFM stack. (a) Input voltage, $V_{IN}$, consisting of a rectangular preset pulse followed by a triangular waveform. (b) Measured current corresponding to the $V_{IN}$ waveform in (a). (c) Total polarization, $P_T$, as a function of the applied electric field. $P_r$ and $E_C$ denote respectively the remnant polarization and the coercive field.}
	\label{Fig:Ferro_Meas_Waves}
\end{figure}
The total polarization $P_T$ in a ferroelectric material can be written as $P_T$$=$$P$$+$$(\varepsilon_{0} \varepsilon_{F}$$-$$1) E_{FE}$, where {\bf $E_{FE}$} is the electric field in the ferroelectric, $\varepsilon_{0}$ and $\varepsilon_{F}$ are respectively the vacuum and background ferroelectric permittivity, and $P$ is the spontaneous polarization.
% that confers to these materials their ferroelectric properties. 
%features of the overall polarization versus field curves, including the hysteretic behavior shown in Fig.\ref{Fig:NC_hyst}. The 
%\begin{equation}
%	P_T = P +  (\varepsilon_{0} \varepsilon_{F} -1) \, F 
%	\label{Eq:Displ}
%\end{equation}
% , and $P$ is assumed to be normal to the ferroelectric interface.
From a theoretical perspective, it has been argued that $\varepsilon_F$ is more an adjustable parameter than a true material constant \cite{Levanyuk_Ferroelectrics2016}.
%, and in practice it is difficult to extract $\varepsilon_F$  independently of the anysotrpy constants that govern the spontaneous polarization (see Sec.\ref{sec:MicroMacroModels}).
In thin film ferroelectrics, moreover, a contribution to $\varepsilon_{F}$ may also stem from the 
%\sout{fact that a mixture of ferroelectric and paraelectric phases is typically observed} {\red 
coexistence of both ferroelectric and paraelectric phases in actual samples.
%Moreover, in thin ferroelectric films we usually observe s that contribute to $\varepsilon_{F}$.
%hence these materials also have a linear polarizability, that is in fact expressed by the background polarization in Eq.\ref{Eq:Displ}.
%
\begin{figure}[h!]
        \centering
        \includegraphics[width = 4in]{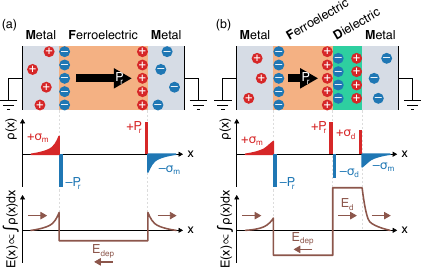}
        \caption{\label{Fig:Depolar_Sketch} \protect Sketch of the depolarization field, $E_{dep}$, induced by the ferroelectric polarization when it is not completely screened by ideal metal electrodes. a) MFM structure where metals have a finite screening length, hence they respond to $P$ with a charge per unit area $\sigma_m$ that is smaller than $P$. b) MFDM structure where $P$ induces $E_{dep}$ in the ferroelectric, as well as the field $E_{d}$ in the dielectric.
        }
\end{figure}

Metal-Ferroelectric-Metal (MFM) capacitors are the primary structures for the electrical characterization of ferroelectric materials.
%Many devices based on ferroelectrics, however, may have additional layers in series to the ferroelectric. The simplest example is the Metal-Ferroelectric-Insulator-Metal (MFIM) structure, that is used, for example, in Ferroelectric Tunnelling Junctions (FTJs). 
The polarization in an MFM stack is typically measured by applying a triangular voltage waveform $V_{IN}$ \cite{Massarotto_SSE23}, as illustrated in \cref{Fig:Ferro_Meas_Waves}(a). 
%\sout{A preset pulse is used before issuing the $V_{IN}$ waveform, so as to set the ferroelectric in the positive polarization state.} {\red Usually, a preset pulse is issued before the triangular waveform in order to set the ferroelectric to a known initial polarization state.}
%; the total polarization is frequently simply as charge, namely $Q$$\simeq$$P_T$.
%The measurements in Fig.\ref{Fig:Ferro_Meas_Waves} were taken in an FTJ consisting of an MFIM structure, but qualitatively similar current and charge plots are obtained for a simpler MFM structure.

\Cref{Fig:Ferro_Meas_Waves}(b) displays an example of the current, $I$, induced by the $V_{IN}$ waveform in \cref{Fig:Ferro_Meas_Waves}(a): the regions where $I$ is fairly flat correspond to the linear dielectric response of the MFM stack, while the $I$ peaks originate from the switching of $P$. Assuming that $I$ has reached a time-periodic regime, the current is integrated over time to obtain the polarization versus field curves in \cref{Fig:Ferro_Meas_Waves}(c).
% charge that is typically interpreted as a measure of the polarization \cref{Fig:Ferro_Meas_Waves}(c).
% the current waveform has readche a priodic behavior 
%, which in effect is the charge variation with respect to the initial charge.
%
%By issuing a few cycles of the $V_{IN}$ waveform shown in Fig.\ref{Fig:Ferro_Meas_Waves}(a), the corresponding current and charge waveforms become also periodic in time, so that it is possible to eliminate the time and plot the polarization versus the ferroelectric field as it is displayed in Fig.\ref{Fig:Ferro_Meas_Waves}(d), where $P_R$ denotes the remnant polarization and $E_C$ the coercive field.

Hafnium-based oxides exhibit $P_r$ values between roughly 5 and \SI{25}{\micro\coulomb\per\centi\metre\squared} depending on the doping \cite{Wang_ScientificReport2021}, while Al$_{1-x}$Sc$_{x}$N can reach \SI{100}{\micro\coulomb\per\centi\metre\squared} \cite{Fichtner_JAP2019}. These figures correspond to huge charges per unit area; in fact, we recall that \SI{16}{\micro\coulomb\per\centi\metre\squared} corresponds to about \SI{e14}{\per\centi\metre\squared} electron charges. Consequently, unless the ferroelectric material is placed between two ideal metals, the polarization tends to induce an electric field in the ferroelectric and in adjacent materials, as it is illustrated in \cref{Fig:Depolar_Sketch}. The field in the ferroelectric is opposite to the polarization, therefore it is called the depolarization field.
%because it tends to destabilize the polarization and ultimately depolarize the material.

%Depolarization fields can induce polarization-dependent built-in potentials, which are indeed the working principle of most ferroelectric-based electron devices (see \cref{sec:Devices}).

%
\begin{figure}
\centering{\phantomsubcaption\label{Fig:Polar_Reversal_waveform}\phantomsubcaption\label{Fig:Polar_Reversal_KAI}\phantomsubcaption\label{Fig:Polar_Reversal_NLS}}
        \includegraphics[width = 4in]{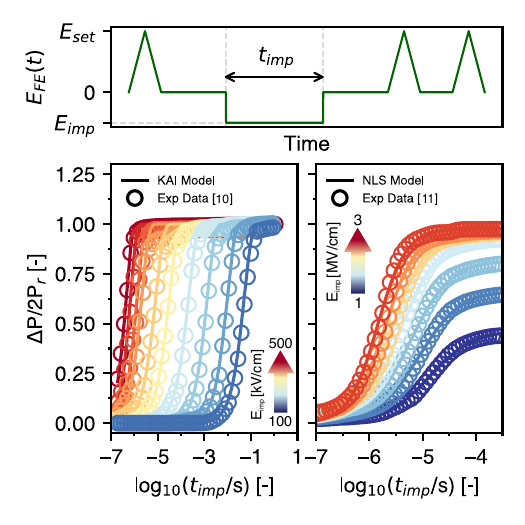}
        \putHelvetica{-25}{175}{12}{a)}
        \putHelvetica{-115}{145}{12}{b)}
        \putHelvetica{-25}{145}{12}{c)}
	\caption{
 %Polarization reversal curves and applied waveform for two different ferroelectric materials. 
 a) Electric Field waveform applied to the MFM stack in order to extract the polarization reversal characteristics. 
 %The sample is first preset at $E_{set}$ and then is kept at a fixed electric field $E_{imp}$ for an amount of time $t_{imp}$. The amount of switched polarization during the pulse is extracted from the total charge with two following electric field pulses of the same magnitude as the preset, as is usually done in Positive-Up-Negative-Down (PUND) measurements. 
 b) Polarization reversal curves for an epitaxial, 10 nm thick perovskite ferroelectric \cite{So_APL2005}. 
 %MFM stack with a thickness of \SI{10}{\nano\meter} reported in \cite{So_APL2005}. 
 c) Polarization reversal curves for a  \SI{9.5}{\nano\meter} thick, poly-cristalline hafnium-zirconium ferroelectric \cite{Lee_ActMat2022}.}
 %MFM stack with a thickness of \SI{9.5}{\nano\meter} reported in \cite{Lee_ActMat2022}.}
 \label{Fig:Polar_Reversal}%
 \vspace{-4mm}
\end{figure}
Besides the polarization versus field curve discussed in \cref{Fig:Ferro_Meas_Waves}(c), the polarization reversal experiments are also very important for the characterization and modelling of ferroelectrics. 
Polarization reversal is usually studied as a function of the applied field and pulse duration in MFM capacitors. As it can be seen in \cref{Fig:Polar_Reversal}, the switching time drastically reduces by increasing the electric field. 
%Polarization reversal is generally assumed to occur through two phases: a) the nucleation in few sites, where the polarization is reversed; b) the growth of such sites and a progressive coalescence of reversed domains. 
%
Two qualitatively different ferroelectric dynamics and corresponding models have been reported in the literature. The Kolmogorov-Avrami-Ishibashi (KAI) model is based on the idea that the kinetics is limited by the rate of domain propagation \cite{Orihara_JPSJ1994}, and it describes well the polarization reversal in epitaxial materials where a single time constant is observed, as in \cref{Fig:Polar_Reversal}(a). The Nucleation Limited Switching (NLS) regime is instead typically used for poly-crystalline materials \cite{Tagantsev_PHRB2002}, where the kinetics has a stretched exponential time dependence, stemming from many different domain nucleation times, as in \cref{Fig:Polar_Reversal}(b).
% , and it is believed to be dominated by the rate of domain nucleation 

%%%%%%%%%%%%%%%%%%%%%%%%%%%%%%%%%%%%%%%%%%%%%%%%%%%%
\section{Microscopic and Macroscopic Models}\label{sec:MicroMacroModels}
%%%%%%%%%%%%%%%%%%%%%%%%%%%%%%%%%%%%%%%%%%%%%%%%%%%%

\begin{figure}[h!]
        \centering
        \includegraphics[width = 4in]{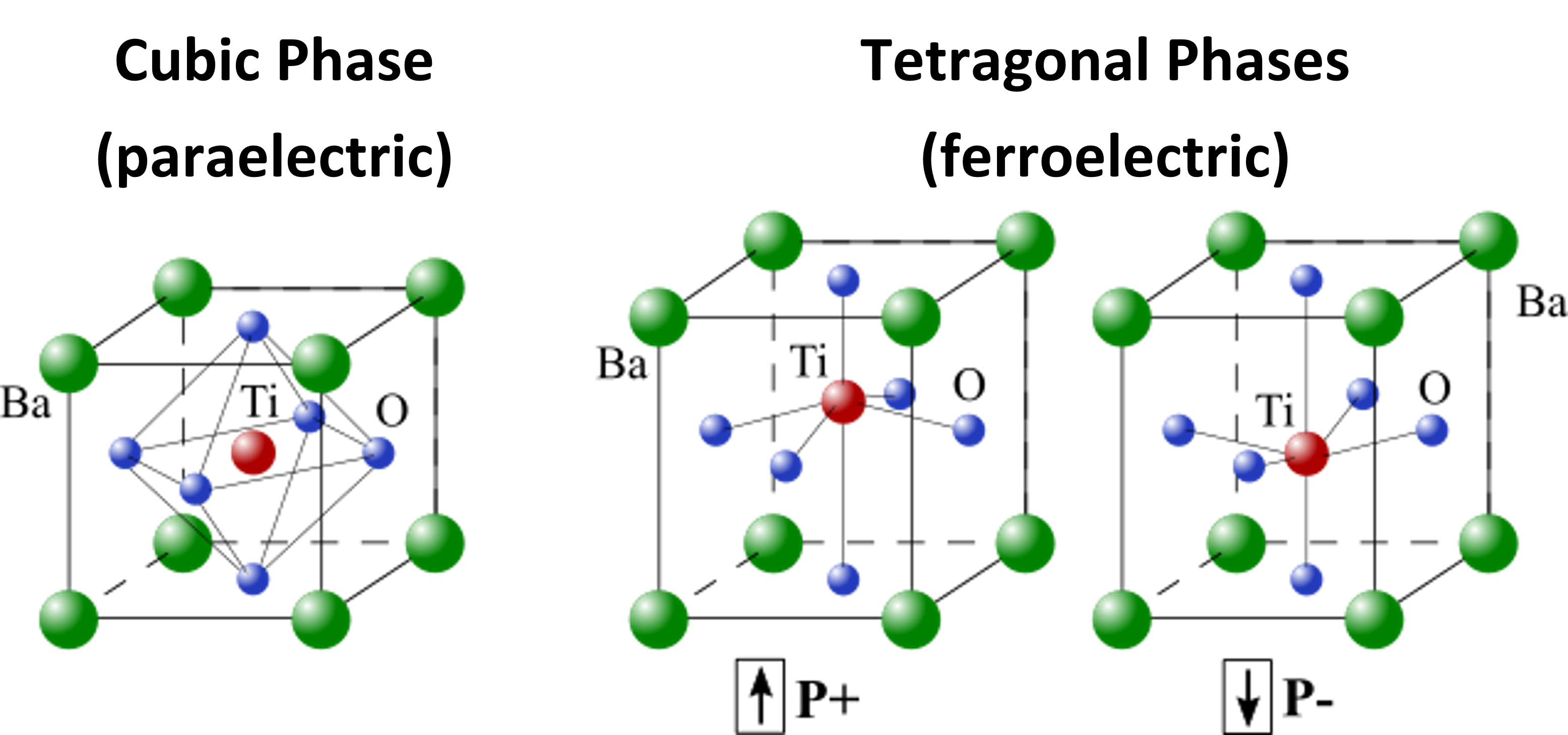}
        \caption{\label{Fig:Sketch_BaTiO} \protect Sketch of the BaTiO$_3$ crystal cell in different phases: the cubic phase does not show any electric momentum and has a paraelectric behavior, while the tetragonal phase can appear in two different configurations with opposite electric momentum, thus showing ferroelectricity. The atom displacements are emphasized to illustrate the two ferroelectric configurations. Adapted from \cite{RabeDawber_Springer2007}.
        }
\end{figure}

Modern physical theories define the polarization in terms of the accumulated adiabatic flow of current occurring when the crystal undergoes a deformation, and it is thus closely related to the Berry phase of the underlying Bloch functions \cite{Resta_Vanderbuilt_Springer2007}. 
% \cite{Resta_Vanderbuilt_Springer2007}. 
%
%{\red This also implies that polarization differences are theoretically more fundamental than absolute polarization values.}
%, which is consistent with how the ferroelectric polarization is measured in electrical characterization experiments (see Fig.\ref{Fig:Ferro_Meas_Waves}, as well as Sec.\ref{sec:PolarizationReversal}).
The concept itself of macroscopic polarization, however, is intuitively linked to electric dipole moments.
%and, in the case of ferroelectric materials, to the presence of spontaneous and switchable microscopic dipoles. 
In a crystalline ferroelectric, the microscopic dipoles are in general due to a lack of inversion symmetry in the unit cell, as it is illustrated in \cref{Fig:Sketch_BaTiO} for BaTiO$_3$.
%, showing that the same material can have both paraelectric and ferroelectric phases that usually coexist in actual samples. 
%
The crystal structure, polarization, dielectric as well as piezoelectric coefficients for perovskites have been long studied by using first-principles methods \cite{Rabe_FirstPrinciples_Springer2007}. More recent studies have been devoted to doped  HfO$_2$ \cite{Qi_PHRB2020},
%Moreover, {\em ab-initio} studies can help identify 
delving also into the most energetically favorable paths for 
%the phase transitions governing the 
polarization reversal \cite{Lee_science2020,CHOE20218}.
%and molecular dynamics simulations can be used to analyze domain-wall motion during polarization switching \cite{Liu_Nature2016}.
%
Moving to device-level modelling, the behavior of spontaneous polarization can be described with a general theory for phase transitions originally proposed by Lev Landau  \cite{Landau_ZETF1937}, that was later applied to ferroelectric materials \cite{Devonshire_PhiMag1949,Landau_Dokl1954}. The model describes the equilibrium and dynamics of polarization in terms of an appropriate thermodynamic potential which, by assuming that the spontaneous polarization $P$ lies in the $z$ direction normal to the ferroelectric interface, we here write in the form \cite{Rollo_Nanoscale2020,Bizindavyi_CommPhysics2021}
\begin{equation}
    \begin{split}
	u_{F}=\alpha P^2 + \beta P^4 + \gamma P^6 &-\dfrac{\varepsilon_0 \varepsilon_{F}}{2}E_{FE}^2 \\
                                                  &- E_{FE} \cdot P + k \, | \nabla P |^2 
    \end{split}
% [J/m^3]
	\label{Eq:UF_LGD}
\end{equation}
where $\alpha$, $\beta$ and $\gamma$ are the ferroelectric anisotropy coefficients, $k$ is the domain wall coupling coefficient and $\nabla P$ is the gradient of $P$. The always-positive, last term in \cref{Eq:UF_LGD} introduces an energy penalty for a non-uniform ferroelectric polarization pattern and, in particular, for configurations with anti-parallel adjacent dipoles. %In other words, this term  tends to couple adjacent domains or elementary dipoles and $k$ is denoted domain wall coupling coefficient. 
%, whose value has been estimated by {\em ab-initio} calculations for some perovkite materials ().

\Cref{Eq:UF_LGD} neglects a possible coupling between polarization and strain,
%, even if ferroelectric materials are usually sensitive to elastic stress and exhibit also piezoelectricity. Such a coupling 
that is frequently overlooked in the analysis of electron devices, but is more relevant for applications to sensors, actuators, and energy harvesting \cite{Chauhan_JAP2017,Li_PHRB2016}.
%Moreover, the influence of stress/strain conditions on ferroelectric properties is itself an instructive research topic 
%
In devices where the ferroelectric is adjacent to dielectrics or semiconductors, the electrostatic energy due to the depolarization field must be included in the thermodynamic potential \cite{Salahuddin_NL2008,Hoffmann_IEDM2018,Rollo_Nanoscale2020, Esseni_Nanoscale2021}. The thermodynamic potential in the presence of conduction of free charges in the ferroelectric has been recently revisited in \cite{Bizindavyi_CommPhysics2021}.
%The Landau–Ginzburg–Devonshire theory (LGD) in Eq.\ref{Eq:LKE} is practically appealing because its phenomenological nature allows one to extract from a comparison to experiments the few parameters governing physical model, namely the anisotropy constants and the domain waal coefficients.
%

The Landau–Ginzburg–Devonshire (LGD)  model 
%Eq.\ref{Eq:LKE} is practically appealing because its phenomenological nature allows one to extract from a comparison to experiments the few parameters governing physical model, namely the anisotropy constants and the domain wall coefficients.
describes the dynamics of $P$ as \cite{Landau_Dokl1954,penrose1990thermodynamically}
%to the variations of the thermodinamic potential, namely we have \cite{Landau_Dokl1954,penrose1990thermodynamically}
\begin{equation}
	\rho \frac{\partial P}{\partial t} = -  \frac{\delta U}{\delta P} =  - \frac{\partial u_F}{\partial P} + 2k \, (\nabla^2 P)
 %\left ( \frac{\partial U}{\partial P}   \right )
	\label{Eq:LGD_Dynamcs}
\end{equation}
where $U$ is the integral of the $u_F$ in \cref{Eq:UF_LGD} over the volume of the system, $\frac{\delta U}{\delta P}$ and $\frac{\partial U}{\partial P}$ respectively denote the variational derivative and the partial derivative of the functional $U$, and $\rho$ is a resistivity (in [\si{\ohm\metre}]) governing the speed of the polarization dynamics. An LGD model for ferroelectrics is also available in TCAD tools \cite{Sentaurus2019,Lenarczyk_SISPAD2016}.
%
%%%%%%%%%%%%%%%%%%%%%%%%%%%%%%%%%%%%%%%%%%%%%%%%%%%%
% \section{Polarization Reversal}\label{sec:PolarizationReversal}
%%%%%%%%%%%%%%%%%%%%%%%%%%%%%%%%%%%%%%%%%%%%%%%%%%%%

As already mentioned, semi-empirical equations are typically used to interpret the polarization reversal experiments in \cref{Fig:Polar_Reversal} \cite{Orihara_JPSJ1994,Tagantsev_PHRB2002}, but an analysis based on the LGD model summarized by \cref{Eq:UF_LGD,Eq:LGD_Dynamcs} could help us test and improve the maturity of the LGD framework.

%%%%%%%%%%%%%%%%%%%%%%%%%%%%%%%%%%%%%%%%%%%%%%%%%%%%
\section{Modelling of Ferroelectric Devices}\label{sec:Devices}
%%%%%%%%%%%%%%%%%%%%%%%%%%%%%%%%%%%%%%%%%%%%%%%%%%%%
%
\begin{figure}
	\centering
	\includegraphics[width=4in]{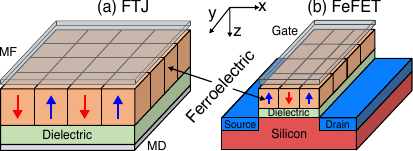}
	\caption[]{Examples of ferroelectric-based electron devices. a) Metal–Ferroelectric–Dielectric–Metal (MFDM) structures have been employed in  Ferroelectric Tunnel Junctions (FTJs) \cite{Max_JEDS2019}.
            % , particularly for a back-end-of-line compatible implementation. 
        b) Ferroelectric FET (FeFETs) can be used either as negative-capacitance FETs \cite{Salahuddin_NL2008,Rollo_EDL2018}, or as memories or memristive devices \cite{Slesazeck_Nanotechnology2019,Mulaosmanovic_EDL2020}.}
	\label{Fig:Fe_Devices}
\end{figure}
The modelling of ferroelectric devices is complicated by the fact that the ferroelectric can be stacked with other dielectrics or semiconductors.
%, as it is illustrated in Fig.\ref{Fig:Fe_Devices}. 
In the MFDM structures employed in FTJs and depicted in Fig.\cref{Fig:Fe_Devices}(a), for example, it has been argued that, if charge injection in the dielectric stack is not accounted for, simulation results predict P-V hysteretic curves much narrower and more tilted than their experimental counterpart \cite{Fontanini_TED2022,Segatto_JEDS2022}. Hence, in these devices, the interplay between the charge trapping, the stabilization and the compensation of the ferroelectric polarization becomes quite delicate \cite{Fontanini_JEDS21}.
The non-hysteretic polarization switching in MFDM structures has been investigated also in the context of the negative-capacitance (NC) operation \cite{Hoffmann_Nature2019,Rollo_Nanoscale2020,Hoffmann_AdvFunctionalMaterials_2021,Esseni_Nanoscale2021}, recently involving also the anti-ferroelectric ZrO$_2$ \cite{Hoffmann_NatComm2022,Segatto_TED2023}.

The ferroelectric FETs (FeFETs) sketched in \cref{Fig:Fe_Devices}(b) have first attracted much attention as steep-slope transistors based on the ferroelectric NC behavior \cite{Salahuddin_NL2008,Rollo_EDL2018}, then they have been investigated for their potentials as memory or memristive devices \cite{Slesazeck_Nanotechnology2019,Mulaosmanovic_EDL2020}. The modelling of FeFETs calls for computationally demanding three-dimensional simulations, because a description of the percolation source-to-drain current paths is essential to study a possible multi-level device operation \cite{Lizzit_ESSDERC2022}. 
% \cite{Lizzit_ESSDERC2021,Lizzit_ESSDERC2022}.

%%%%%%%%%%%%%%%%%%%%%%%%%%%%%%%%%%%%%%%%%%%%%%%%%%%%
\section{Outlook and conclusions}\label{sec:conclusion}
%%%%%%%%%%%%%%%%%%%%%%%%%%%%%%%%%%%%%%%%%%%%%%%%%%%%

Ferroelectrics are CMOS-compatible active oxides with a broad set of possible applications. The polarization switching is a field-driven process that is inherently energy efficient, which is of utmost importance for memory and memristive applications.
Some fundamental aspects behind the polarization reversal remain to be understood in fluorite-type Hf$_{0.5}$Zr$_{0.5}$O$_2$ and wurtzite-type Al$_{1-x}$Sc$_{x}$N materials, and this plays a crucial role for the possibility of achieving a multi-level, quasi-analog operation of ferroelectric devices. 
Many challenges and opportunities still exist for the modelling of ferroelectric materials at different abstraction levels, and the modelling will play an important role in order to harness the potentials of ferroelectric materials in nanoelectronic devices and circuits.

\section*{Acknowledgments}
%{\bf Acknowledg.}
This work was supported by the European Union through the BeFerroSynaptic project (GA:871737).

%%%%%%%%%%%%%%%%%%%%%%%%%%%%%%%%%%%%%%%%%%%%%% SISPAD 2023 BIBLIOGRAPHY %%%%%%%%%%%%%%%%%%%%%%%%%%%%%%%%%%%%%%%%%
% BIBTEX BIBLIOGRAPHY
\bibliographystyle{IEEEtran}
\bibliography{bibliography}
    
\end{document}